\documentclass{IEEEtran}
\usepackage{cite}
\usepackage{amsmath,amssymb,amsfonts}
\usepackage{algorithmic}
\usepackage{graphicx}
\usepackage{textcomp}
\usepackage{epstopdf}
\def\BibTeX{{\rm B\kern-.05em{\sc i\kern-.025em b}\kern-.08em
    T\kern-.1667em\lower.7ex\hbox{E}\kern-.125emX}}
\begin{document}
\title{An Accurate Edge-based FEM for Electromagnetic Analysis with Its Applications to \\ Multiscale Structures}
\author{Yangfan Zhang, \IEEEmembership{Graduated Student Member, IEEE}, Pengfei Wang, Weiping Li, \\ and Shunchuan Yang, \IEEEmembership{Member, IEEE}
\thanks{This work was supported by the Beijing Natural Science Foundation under Grant 4194082, National Natural Science Foundation of China through Grant 61801010, and Fundamental Research Funds for the Central Universities. {\it{ (Corresponding author: Shunchuan Yang)}}

Yangfan Zhang and Shunchuan Yang are with the School of Electronic and Information Engineering, Beihang University, Beijing, 100191, CN (e-mail: yfz2018@buaa.edu.cn, scyang@buaa.edu.cn). 

Pengfei Wang is with the School of Instrumentation Science and Opto-electronics Engineering, Beihang University, Beijing, 100191, CN (e-mail: wangpengfei2013@buaa.edu.cn)

Weiping Li are with the School of Physics, Beihang University, Beijing, 100191, CN (e-mail: liwp@buaa.edu.cn).}}

\maketitle

\begin{abstract}
This paper introduces an accurate edge-based smoothed finite element method (ES-FEM) for electromagnetic analysis for both two dimensional cylindrical and three dimensional cartesian systems, which shows much better performance in terms of accuracy and numerical stability for mesh distortion compared with the traditional FEM. Unlike the traditional FEM, the computational domain in ES-FEM is divided into nonoverlapping smoothing domains associated with each edge of elements, triangles in two dimensional domain and tetrahedrons in three dimensional domain. Then, the gradient smoothing technique (GST) is used to smooth the gradient components in the stiff matrix of the FEM. Several numerical experiments are carried out to validate its accuracy and numerical stability. Numerical results show that the ES-FEM can obtain much more accurate results and is almost independent of mesh distortion.
\end{abstract}


\section{Introduction}
\label{sec:introduction}
The finite element method (FEM) is one of the most powerful numerical methods to solve practical engineering problems due to its strong capability of handling complex and multiscale structures \cite{ref1}. In the computational electromagnetics (CEM), the FEM is widely used to solve various electromagnetic problems, like electrostatic currents [2,3], magneticstatic problems \cite{ref4}, electromagnetic scattering \cite{ref5}, integrated circuits modeling \cite{ref6}, electrical-thermal co-simulations \cite{ref7}. 

However, investigations found that the traditional FEM is quite sensitive to the quality of background meshes \cite{ref1}. More specifically, equilateral triangles and regular tetrahedrons are much preferred to guarantee its accuracy. When complex or multiscale structures are discretized into many nonoverlapping elements, a large number of irregular elements may exist. Then, the accuracy of the FEM can be severely degenerated and even totally unacceptable. However, generation of high good quality triangles and tetrahedrons is quite challenging when multiscale and complex structures are involved \cite{ref8}. Therefore, the results cannot be always trusted when the mesh quality cannot be guaranteed. In addition, when low order basis functions are used, only a low level of accuracy of numerical results can be obtained. That implies that a large number of unknowns are required if highly accurate results are needed \cite{ref9}.

In this paper, we address these two issues in the CEM by introducing the edge-based smoothed FEM (ES-FEM) to accurately solve the electromagnetic problems in the two dimensional cylindrical and three dimensional cartesian systems. The ES-FEM was firstly introduced to solve non-electromagnetic problems by properly combining the traditional FEM and mesh-free methods, termed as smoothed point interpolation methods (S-PIMs)\cite{ref10}. Various versions of smoothed FEM (SFEM) are proposed based on how to construct smoothing domains, like node-based FEM (NS-FEM) \cite{ref11}, face-based FEM (FS-FEM) \cite{ref12}, edge-based FEM (ES-FEM) \cite{ref13}. They all show better performance in terms of accuracy and numerical stability in thermal analysis \cite{ref14}, acoustic analysis \cite{ref15} and computational mechanics \cite{ref16}. In most of those applications, the ES-FEM shows best performance. In \cite{ref17}, the FS-PIM is introduced for the electromagnetic analysis. It is found that the FS-PIM shows improved accuracy. We introduced the ES-FEM to model high-speed interconnects \cite{nemo} and electrostatic lens \cite{apwc}, which shows much better accuracy and mesh stability compared with the FEM. In this paper, a new two dimensional cylindrical and full three dimensional ES-FEM is proposed for electromagnetic analysis and comprehensively investigated upon its numerical properties, which has significant performance improvement without increasing the number of unknowns.

This paper is organized as follows. In Section II, detailed formulations and its numerical implementations of the ES-FEM are presented. In Section III, its accuracy and numerical properties are comprehensively investigated through several numerical experiments. At last, we draw some conclusions in Section IV.

\section{Formulations}
\subsection{Problem Configurations}
There are various axis-symmetrical systems, like transmission lines \cite{ref2}, electrostatic lens \cite{apwc}, horn antenna \cite{ref20}, which can be simplified as two dimensional problems in the cylindrical system. The boundary-value problems are then defined by the following partial differential equation (PDE) as
%-----------------------------------------------------------------------
% Equation.1
%-----------------------------------------------------------------------
\begin{equation}{\label{D2}}
\frac{{{\partial ^2}V}}{{\partial {r^2}}} + \frac{1}{r}\frac{{\partial V}}{{\partial r}} + \frac{{{\partial ^2}V}}{{\partial {z^2}}} = f,
\end{equation}
subject to the boundary condition
%-----------------------------------------------------------------------
% Equation.2
%-----------------------------------------------------------------------
\begin{equation}{\label{BC1}}
a \frac{{\partial V}}{{\partial n}} + \gamma V = q,
\end{equation}
where $V$ denotes potential in the computational domain,  $f$ is the excitation function, $a$ and $\gamma$ are constants for the third kind of boundary conditions, Equ. (\ref{BC1}) is the first kind of boundary condition with $a = 0$ and $\gamma \ne 0$, and the second kind of boundary condition with $a \ne 0$ and $\gamma = 0$.

For those full three dimensional structures, the PDE \cite{ref1} can be defined as
%-----------------------------------------------------------------------
% Equation.3
%-----------------------------------------------------------------------
\begin{equation}{\label{D3}}
- \frac{\partial }{{\partial x}}\left( {{\alpha _x}\frac{{\partial V}}{{\partial x}}} \right) - \frac{\partial }{{\partial y}}\left( {{\alpha _y}\frac{{\partial V}}{{\partial y}}} \right) - \frac{\partial }{{\partial z}}\left( {{\alpha _z}\frac{{\partial V}}{{\partial z}}} \right) + \beta V = f,
\end{equation}
subject to the boundary condition
%-----------------------------------------------------------------------
% Equation.4
%-----------------------------------------------------------------------
\begin{equation}
a \frac{{\partial V}}{{\partial n}} + \gamma V = q,
\end{equation}
where {$\alpha _x$}, {$\alpha _y$}, and {$\alpha _z$} are constant material parameters, {$\beta$} is a position-dependent function which is associated with  physical properties of the domain, $V$ denotes the potential,  {$f$} is the excitation function, $a$ and $\gamma$ are constants. 

\subsection{Formulations of the FEM}
In the FEM, we can formulate Equ. (\ref{D2}) and Equ. (\ref{D3}) into a set of linear equations with the expansion of $V$ through linear basis functions in each element \cite{ref1}. In its matrix form, those linear equations can be rewritten into 
%-----------------------------------------------------------------------
% Equation.5
%-----------------------------------------------------------------------
\begin{equation}{\label{galerkin}}
\bf{K} \bf{\phi }  -  \bf{b}  = 0,
\end{equation}
where $\bf K$ is assembled from its elemental entity $K^e$ in cylindrical system given by
%-----------------------------------------------------------------------
% Equation.6
%-----------------------------------------------------------------------
\begin{equation}{\label{KE_cy}}
K^{e}=2 \pi \int_{\Omega} \nabla N^{e} \cdot\left(\nabla N^{e}\right)^{T} r d r d z.
\end{equation}
In three dimensional cartesian system,  $K^e$ is expressed as
%-----------------------------------------------------------------------
% Equation.7
%-----------------------------------------------------------------------
\begin{equation}{\label{KE_ca}}
{K^e} = \int_\Omega  {\nabla {N^e} \cdot {{(\nabla {N^e})}^T}} d\Omega ,
\end{equation}
where $N^e$ is a linear shape function in the FEM. 

We will present how to construct the ES-FEM to solve Equ. \eqref{D2} and Equ. \eqref{D3} based on the traditional FEM. In the next subsection, we first introduce the smoothing domain in ES-FEM based on the background mesh.

\subsection{Creation of Smoothing Domain}
Assuming that the solution domain $\Omega$ has been divided into $N_e$ nonoverlapping elements with $N_n$ nodes and $N_{eg}$ edges. For two dimensional domain, by connecting the centroid of each element with its neighbor nodes, as shown in Fig. 1(a), the smoothing domain $\Omega_k^s$ associated with the $k$th edge is created. Therefore, the whole computational domain $\Omega$ is divided into $N_{eg}$ nonoverlapping smoothing domains. Each smoothing domain may consist of 3 or 4 segments and nodes depending on edge location.

% Fig.1
%----------------------------------------------------------------------
\begin{figure}[!t]
	\begin{minipage}[h]{0.48\linewidth}\label{FIG1A}
		\centerline{\includegraphics[width=1.81in]{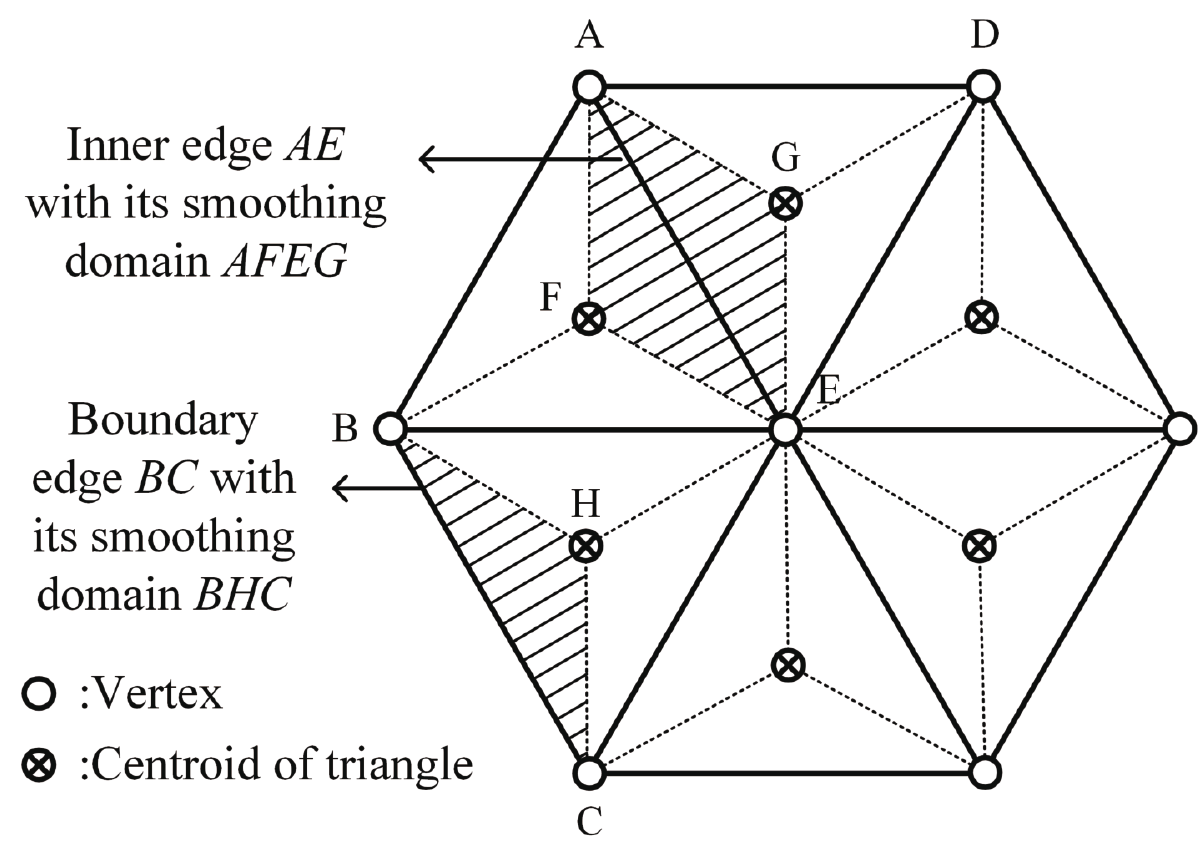}}
		\centerline{(a)}
	\end{minipage}
	\hfill
	\begin{minipage}[h]{0.48\linewidth}\label{FIG1B}
		\centerline{\includegraphics[width=1.81in]{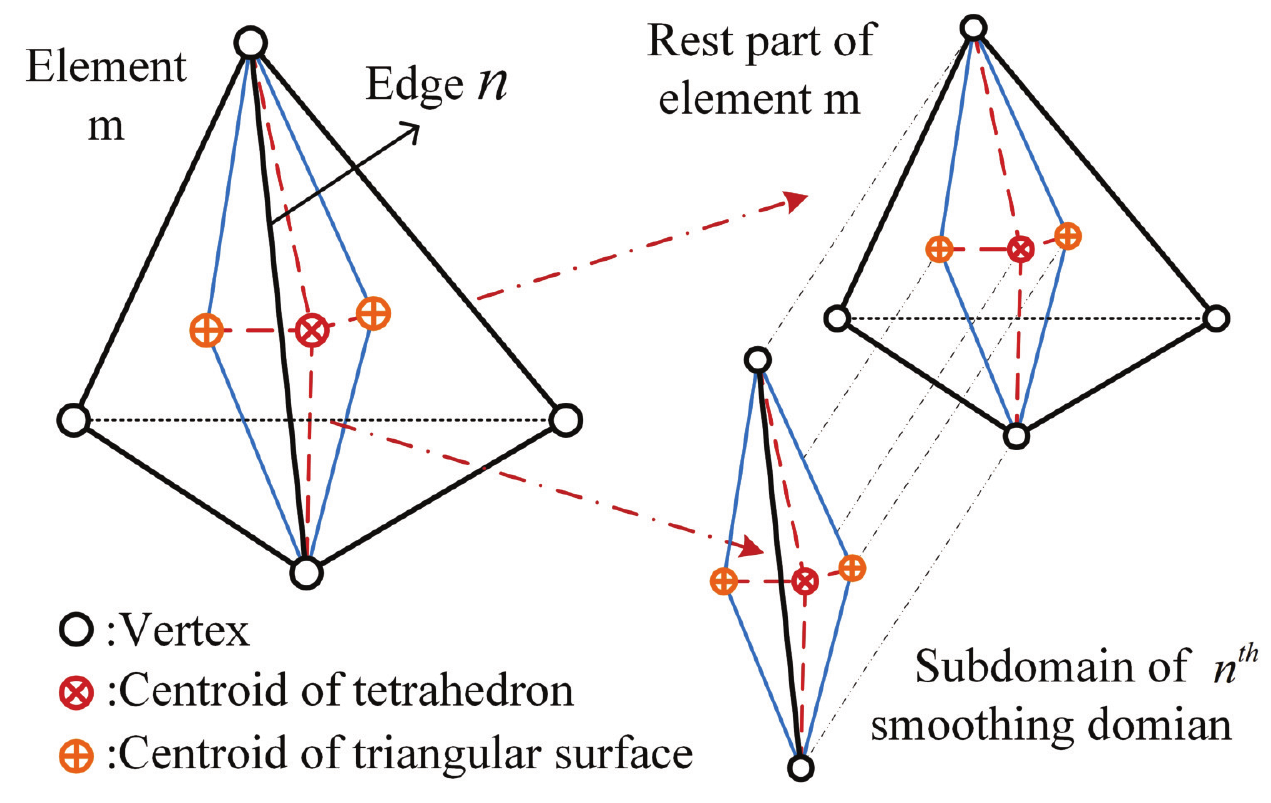}}
		\centerline{(b)}
	\end{minipage}
		\caption{Smoothing domains in different systems: (a) two dimensional case, and (b) three dimensional case.}
	\label{fig_1}
\end{figure}
%----------------------------------------------------------------------

For three dimensional domain, one of the sub-domain of the smoothing domain $\Omega_k^n$ for $n$th edge located in the $m$th tetrahedron is created by connecting the centroid of the tetrahedron element and two centroids of the triangle surfaces, which are shared by edge $n$ and its vertexes, as shown in Fig. 1(b). By assembling all sub-domains of $n$th edge, the $n$th smoothing domain is created.

In traditional FEM, each node has only interactions with those belonging to the same element. However, in the ES-FEM, each node interacts with those contributing to the same smoothing domain. That means the interactions are performed on the smoothing domain rather than triangle or tetrahedron elements, which makes the ES-FEM more accurate and more stable to irregular meshes, especially, for multiscale structures without increasing the overall count of unknowns compared with the traditional FEM.

\subsection{Detailed Formulations of the ES-FEM to Solve EM Problems}
The difference between the ES-FEM and the FEM lies in how the stiff matrix is constructed which contains the gradient of the shape function $\nabla N$. We  replace the gradient component $\nabla N^e$ by its smoothed counterpart $\overline {\nabla N^e}$ through the generalized gradient smoothing technique (GST) \cite{ref10}. In the ES-FEM formulation, the smoothed $\overline {K_d}$ in two dimensional system can be expressed as 
%-----------------------------------------------------------------------
% Equation.8.1
%-----------------------------------------------------------------------
\begin{equation}{\label{Kd1}}
{\overline {{K_d}} ^e} = 2\pi \int_\Omega  {{{({{\overline {\nabla N} }^e})}^T}}  \cdot {\overline {\nabla N} ^e}rdrdz.
\end{equation}
%-----------------------------------------------------------------------

In three dimensional cartesian system, $\overline {K_d}$ is rewritten as

% Equation.8.2
%-----------------------------------------------------------------------
\begin{equation}{\label{Kd2}}
{\overline {{K_d}} ^e} = \int_\Omega  {{{({{\overline {\nabla N} }^e})}^T}}  \cdot {\overline {\nabla N} ^e}d\Omega 
\end{equation}

The GST is applied in each smoothing domain by averaging the gradient shape function 
%-----------------------------------------------------------------------
% Equation.9
%-----------------------------------------------------------------------
\begin{equation}{\label{KD}}
\overline {\nabla N^e} = \int_{\Omega _k^s} {\nabla N^e w\left( {x - {x_k}} \right)} d\Omega,
\end{equation}
where $w\left( {x - {x_k}} \right)$ denotes the smoothing function defined as
%-----------------------------------------------------------------------
% Equation.10
%-----------------------------------------------------------------------
\begin{equation}{\label{W}}
w\left( {x - {x_k}} \right) = \left\{ {\begin{array}{*{20}{c}}
{1/{S_k}}&{x \in \Omega _k^s}\\
0&{x \notin \Omega _k^s}
\end{array}} \right.,
\end{equation}
where $S_k$ is the area or volume of the $k$th smoothing domain in two or three dimensional domains, respectively. Through the divergence theorem [xxx], the gradient smoothed  $\overline {\nabla N^e}$ can be expressed as
%-----------------------------------------------------------------------
% Equation.11
%-----------------------------------------------------------------------
\begin{equation}{\label{SMOOTHEDN}}
\begin{split}
\overline {\nabla N^e} & = \int_\Omega  {\nabla N^e  w ( {x - {x_k}} ) d\Omega } \\
                     &=  - \int_\Omega  {\nabla w\left( {x - {x_k}} \right)N^e d\Omega }  + \int_{\Gamma _k}  {{\vec{ \bf{n}}}N^e w\left( {x - {x_k}} \right)d{\Gamma _k} } \\
                     &= \int_{\Gamma _k}  {{\vec{\bf{n}}}N^e w\left( {x - {x_k}} \right)d\Gamma_k }
\end{split}
\end{equation}
where ${\Gamma_k}$ is the boundary of smoothing domain $\Omega _k^n$, $\vec{\bf{n}}$ is the unit normal vector pointing to outside of the smoothing domain on ${\Gamma_k}$.

When the divergence theorem is applied, we assumed that potential are continuous on the interface of elements. In the eletrostatic problem, this implication is automatically satisfied, since potential is continuous in the computational domain. However, when it comes to vector electric fields, only tangential components are continuous, and normal components on the boundary of media are discontinuous. Carefully attention should be paid to handling this boundary condition.Currently, extension of current work to solve full vector wave equations is in progress.

Equ. (\ref{SMOOTHEDN}) can be easy to be evaluated through Gaussian quadrature numerical integration as
%-----------------------------------------------------------------------
% Equation.12
%-----------------------------------------------------------------------
\begin{equation}{\label{b}}
(\overline {\nabla N})_d = \frac{1}{{{S_k}}}\int_{{\Gamma _k}} {{N}{n_d}d} {\Gamma _k} = \frac{1}{{{S_k}}}\sum\limits_{i = 1}^{{N_b}} w_i {{N}\left( {x_i} \right){n_{id}}{l_i}},
\end{equation}
where $d = r,z$ in two dimensional cylindrical domain and $d = x,y, z$ in three dimensional domain, $N_b$ is the segment count of smoothing domain, $x_i$ and $w_i$ are abscissas and weight for Gaussian numerical integration, respectively, ${n_{id}}$ is the $i$th component of unit outward vector, ${l_i}$ is the length of the $i$th segment. Therefore, we can express the smoothed gradient component of shape functions as

%-----------------------------------------------------------------------
% Equation.13
%-----------------------------------------------------------------------
\begin{equation}{\label{InterpolationN}}
\overline {\nabla N({x_k})}  = \sum\limits_{j = 1}^{{n_k}} {\overline {{B_j}} ({x_k})},
\end{equation}
where $n_k$ is the number of nodes which contributs to the $k$th smoothing domain and $\overline B$ is the smoothed gradient matrix elements which can expressed as
%-----------------------------------------------------------------------
% Equation.14
%-----------------------------------------------------------------------
\begin{equation}{\label{smoothedgradientMatrix2}}
{\overline {{B_i}} ^T}({x_k}) = \frac{1}{{S_k^s}}\sum\limits_{i = 1}^{N_e^k} {\frac{1}{3}S_e^iN_e^i},
\end{equation}
and
%-----------------------------------------------------------------------
% Equation.15
%-----------------------------------------------------------------------
\begin{equation}{\label{smoothedgradientMatrix3}}
{\overline {{B_i}} ^T}({x_k}) = \frac{1}{{S_k^s}}\sum\limits_{i = 1}^{N_e^k} {\frac{1}{6}S_e^iN_e^i}, 
\end{equation}
in two and three dimensional systems, respectively, $N_e^n$ is the number of elements that share the edge $n$, $S_e^i$ is the area or volume of the $m$th triangle or tetrahedron element.

By substituting Equ. \eqref{smoothedgradientMatrix2} to Equ. \eqref{Kd1}, elements of matrix  $\overline {K_d}^e$ in two dimensional system are given by

\begin{equation}{\label{Kdeij}}
{\overline {{K_d}} ^e} = \frac{{\pi \overline r^e S_k^s}}{{18}}\left( {{b_i}{b_j} + {c_i}{c_j}} \right),
\end{equation}
In three dimensional system, by substituting Equ. \eqref{smoothedgradientMatrix3} to Equ. \eqref{Kd2}, elements of matrix  $\overline {K_d}^e$ is expressed as
\begin{equation}{\label{Kdeij2}}
{\overline {{K_d}} _{ij}} = \frac{{S_k^s}}{{{\rm{36}}}}\left( {{b_i}{b_j} + {c_i}{c_j}} \right),
\end{equation}
where $\overline {r^e}=(r_1^e+r_2^e+r_3^e)/3$, $i,j=1,2,...N_e^k$.

\subsection{Boundary Condition}
By applying the boundary conditions exactly the same as those in the traditional FEM, we can accurately solve the electromagnetic problems through the ES-FEM.

\section{Numerical Results and Discussion}

\subsection{A Cubical Box}
To analytically investigate the numerical properties of the ES-FEM, we first consider a cubical box as shown in Fig. 2.
% Fig.2
%----------------------------------------------------------
\begin{figure}[!t]
	\centerline{\includegraphics[scale=0.2]{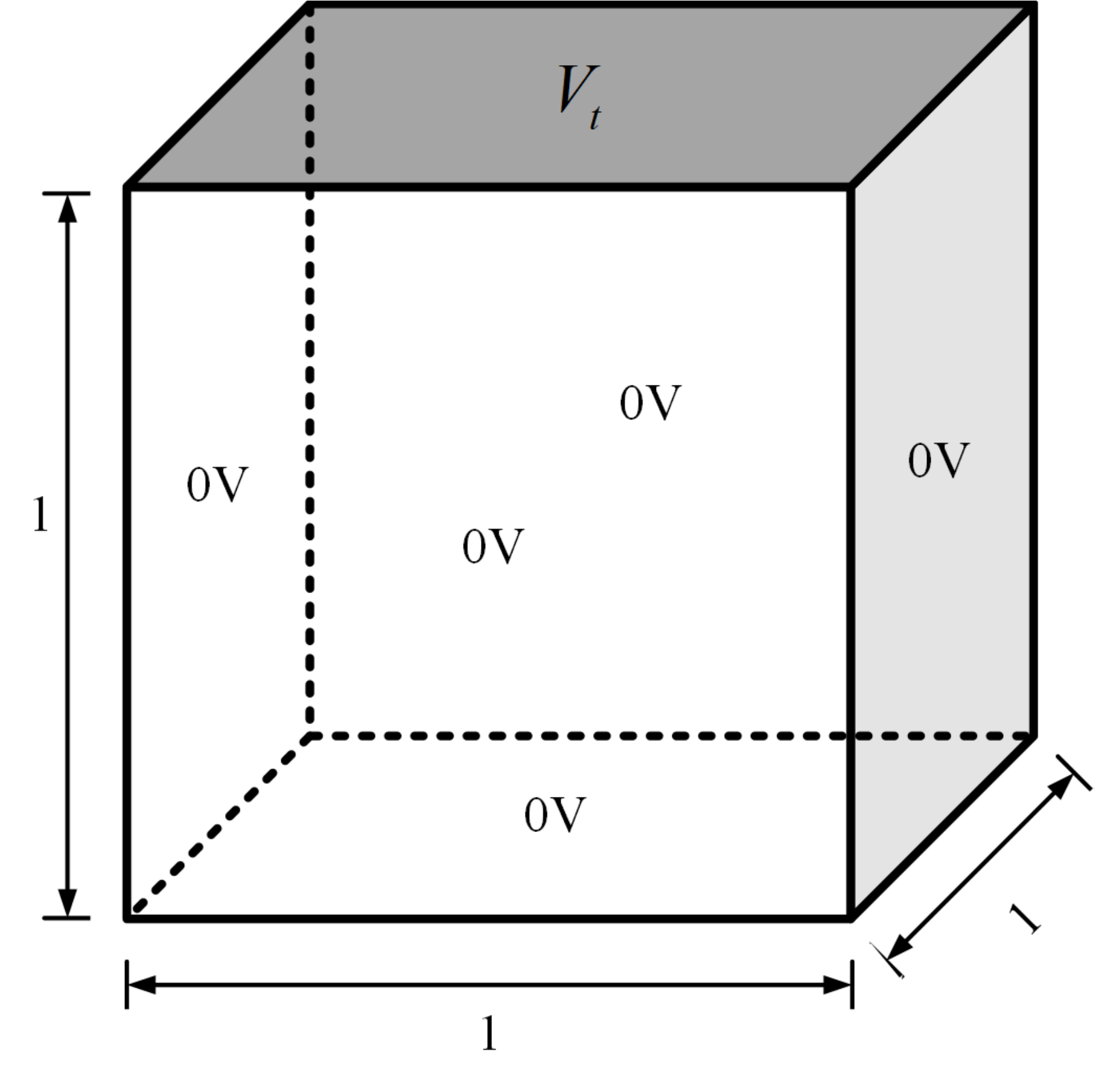}}
	\caption{Geometry configuration of the cubical box and boundary conditions.}
	\label{fig_2}
\end{figure}
%----------------------------------------------------------
The top surface of cubical box is applied to a boundary condition 
\begin{equation}{\label{Vt}}
{V_{t}=10 \sin (\pi x) \sin (\pi y)},
\end{equation}
and other surfaces are set to 0 $V$. Therefore, the potential inside the box is given by

\begin{equation}{\label{Vt}}
{V_{ref}=\frac{10 \sin (\pi x) \sin (\pi y) \sinh (z \pi \sqrt{2})}{\sinh (\pi \sqrt{2})}} .
\end{equation}

% Fig.3
%----------------------------------------------------------
\begin{figure}[!t]
	\centerline{\includegraphics[scale=0.30]{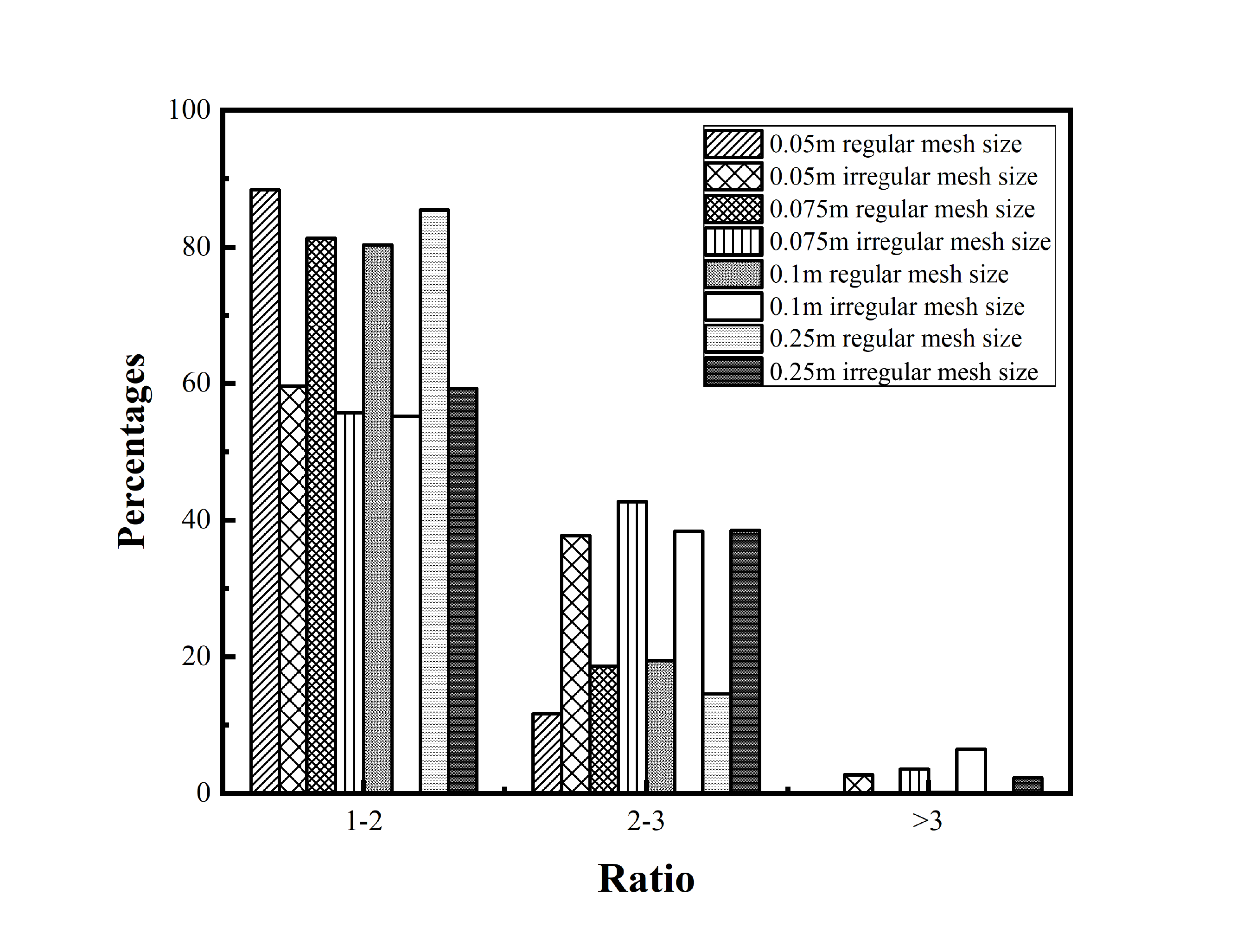}}
	\caption{Ratio distribution in four meshes.}
	\label{fig_3}
\end{figure}
%----------------------------------------------------------

% Fig.4
%----------------------------------------------------------
\begin{figure}[!t]
	\centerline{\includegraphics[scale=0.30]{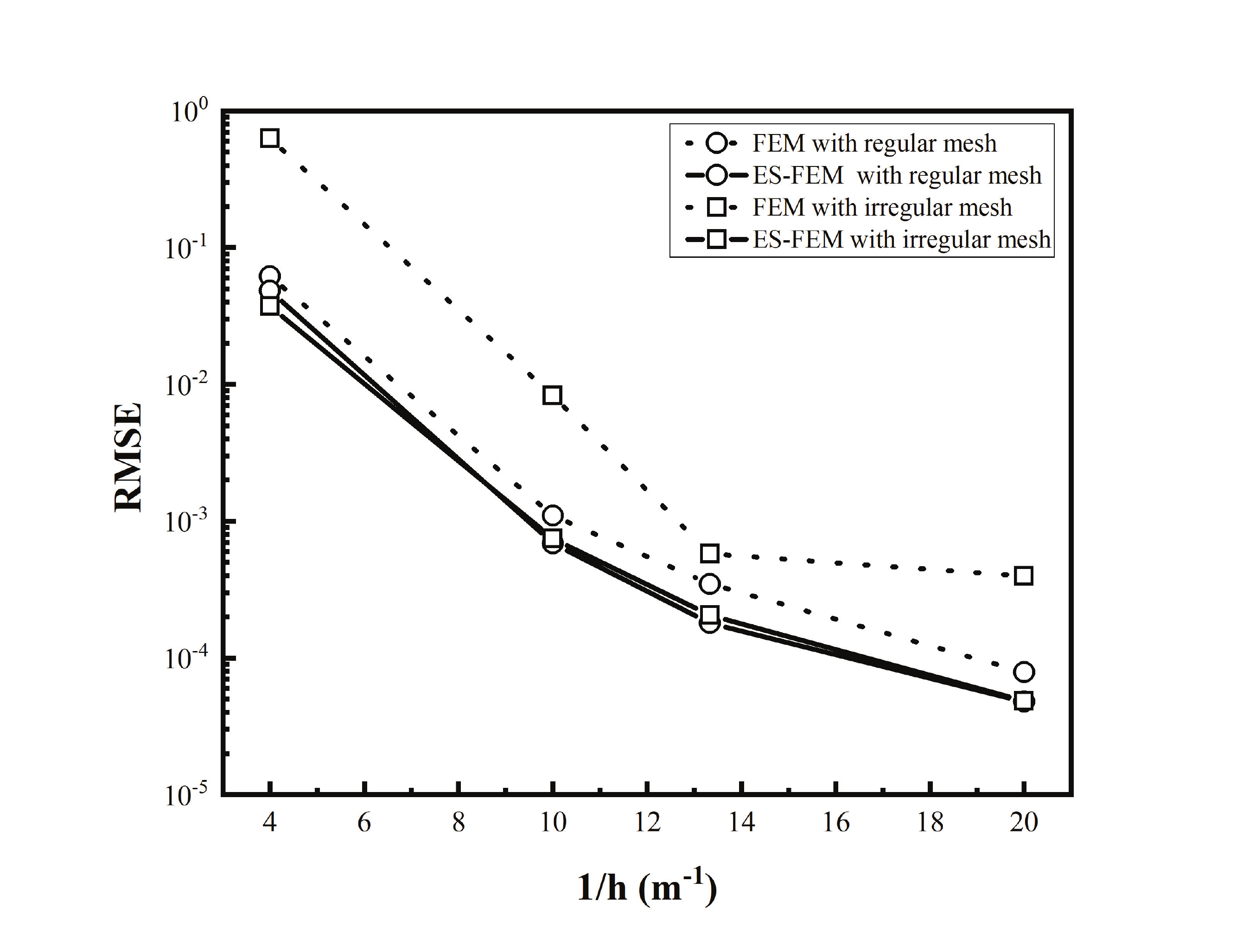}}
	\caption{Relative error obtained from the ES-FEM and the FEM with different meshes.}
	\label{fig_4}
\end{figure}
%----------------------------------------------------------
Four background meshes with different mesh sizes are considered in our numerical analysis. In addition, to investigate the numerical stability of the ES-FEM, four irregular meshes, which are generated through enforcing a small random offset at each node except boundary nodes in the regular meshes, are used. The averaged  edge length of all elements is denoted as {$h$}. The root mean square error (RMSE) of $V$ at all the nodes in the computational domain is defined as 
\begin{equation}{\label{RelativeError}}
{RMSE} = \sqrt {\frac{{\sum\limits_{i = 1}^n {{{\left( {V_s^i - V_{ref}^i} \right)}^2}} }}{{\sum\limits_{i = 1}^n {(V{_{ref}^i})^2}} }} ,
\end{equation}
where ${V_s^i}$ is the numerical result and ${V_{ref}^i}$ is the reference potential obtained from Equ. (\ref{Vt}).

Fig. 3 quantitively illustrates the percentage of different ratio of elements in the four regular and irregular meshes used in the FEM and the ES-FEM. The ratio is defined as the maximum to minimum edge length in each element, which denotes mesh quality. Obviously, the ratio should be as near as one for good quality meshes. As shown in Fig. 3, the ratio of only around 20\% elements is larger than two. However, in irregular meshes, the ratio of over 40\% elements is larger than two, which implies that more distorted elements exist in the irregular meshes. 

Fig. 4 shows RMSE of the traditional FEM and the ES-FEM. It is easy to find that the ES-FEM outperforms the traditional FEM in both regular and irregular meshes. For regular meshes, the ES-FEM is more accurate than the FEM. When it comes to irregular meshes, this method becomes more obvious. As shown in Fig. 4, the accuracy of the FEM is severely degenerated up to almost one order. However, the ES-FEM shows great numerical stability and its accuracy is not affected by the irregular meshes, which shows high level of accuracy and great numerical stability.

\subsection{A Multiscale Electronic Lens}
A multiscale electronic lens, which is widely used in the industrial field is considered which consists of an emitter, a suppression electrode, an absorbing electrode and an anode, as is shown in Fig. 5(a). Detailed geometry configurations are demonstrated in Fig. 5(b). The size of the structure is 12 mm, but the size of the emitter tip is only $3 \times {10^{ - 4}}$ mm. The ratio of the size of the electronic lens to the size of the emitter tip reaches $10^5$, which implies that the electronic lens is a typical multiscale structure. 

\begin{figure}[!t]
	\begin{minipage}[h]{0.48\linewidth}\label{FIG5A}
		\centerline{\includegraphics[width=1.95in]{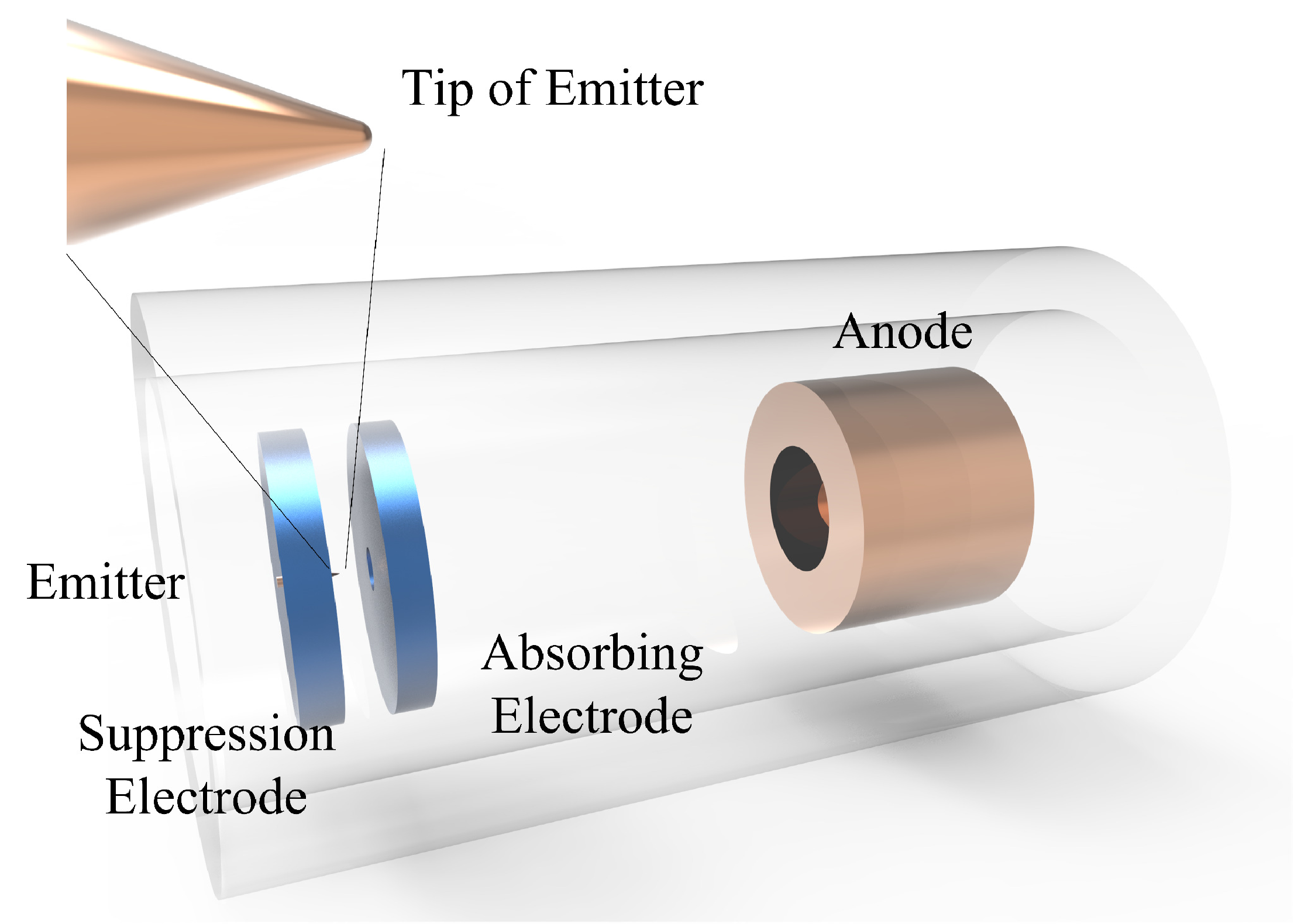}}
		\centerline{(a)}
	\end{minipage}
	\hfill
	\begin{minipage}[h]{0.48\linewidth}\label{FIG5B}
		\centerline{\includegraphics[width=1.95in]{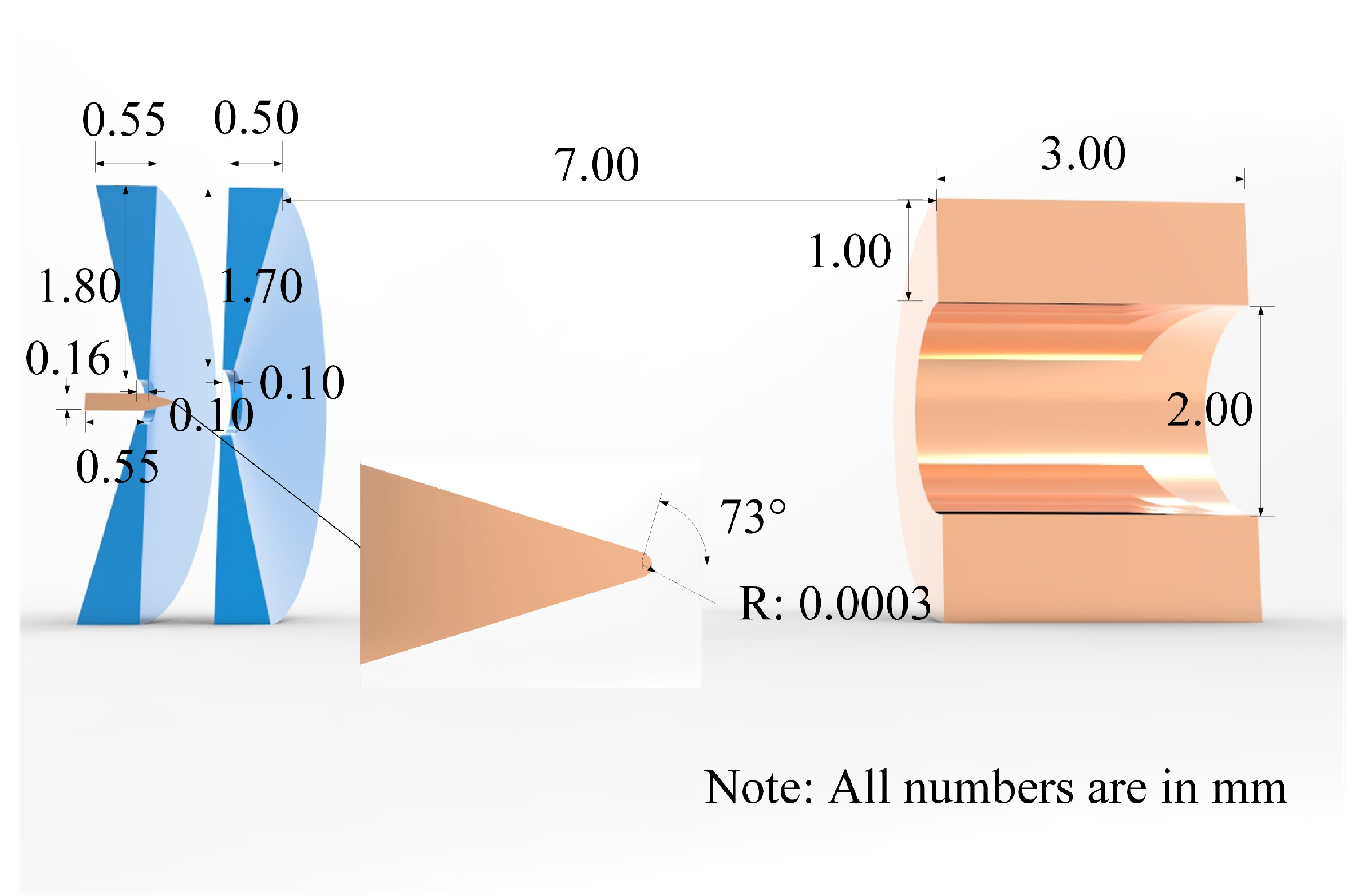}}
		\centerline{(b)}
	\end{minipage}
	\caption{Geometry configurations of the electronic lens.}
	\label{fig_5}
\end{figure}

As shown in Fig. \ref {fig_5}(a), an electronic lens is an axis-symmetric structure which can be simplified into a two dimensional model. The potential of the emitter, the absorbing electrode and the anodes are set to 0 V, 7,000 V and 70,000 V, respectively. Both the two dimensional cylindrical ES-FEM and full three dimensional ES-FEM are applied in this case. Since the potential on the axis is quite important for the electrostatic lens design, we extract the potential on the central axis to verify the accuracy and numerical stability of FEM and ES-FEM. 

Fig. 6 shows that the ratio of elements in the four meshes used in our simulations. As shown in Fig. 6, the ratio of almost 40\% elements is larger than two. Further investigations find that most irregular elements exist near the tip of the emitter. Since the lens is a multiscale structure, the quality of meshes cannot be improved no matter how we refine the meshes. It is challenging to generate good quality meshes for multiscale structures. Since the FEM is sensitive to mesh quality, the accuracy cannot be guaranteed for multiscale structures \cite{ref9}.
% Fig.5
%----------------------------------------------------------------------
%---------------------------------------------------------------------

% Fig.6
%----------------------------------------------------------
\begin{figure}[!t]
\centerline{\includegraphics[scale=0.30]{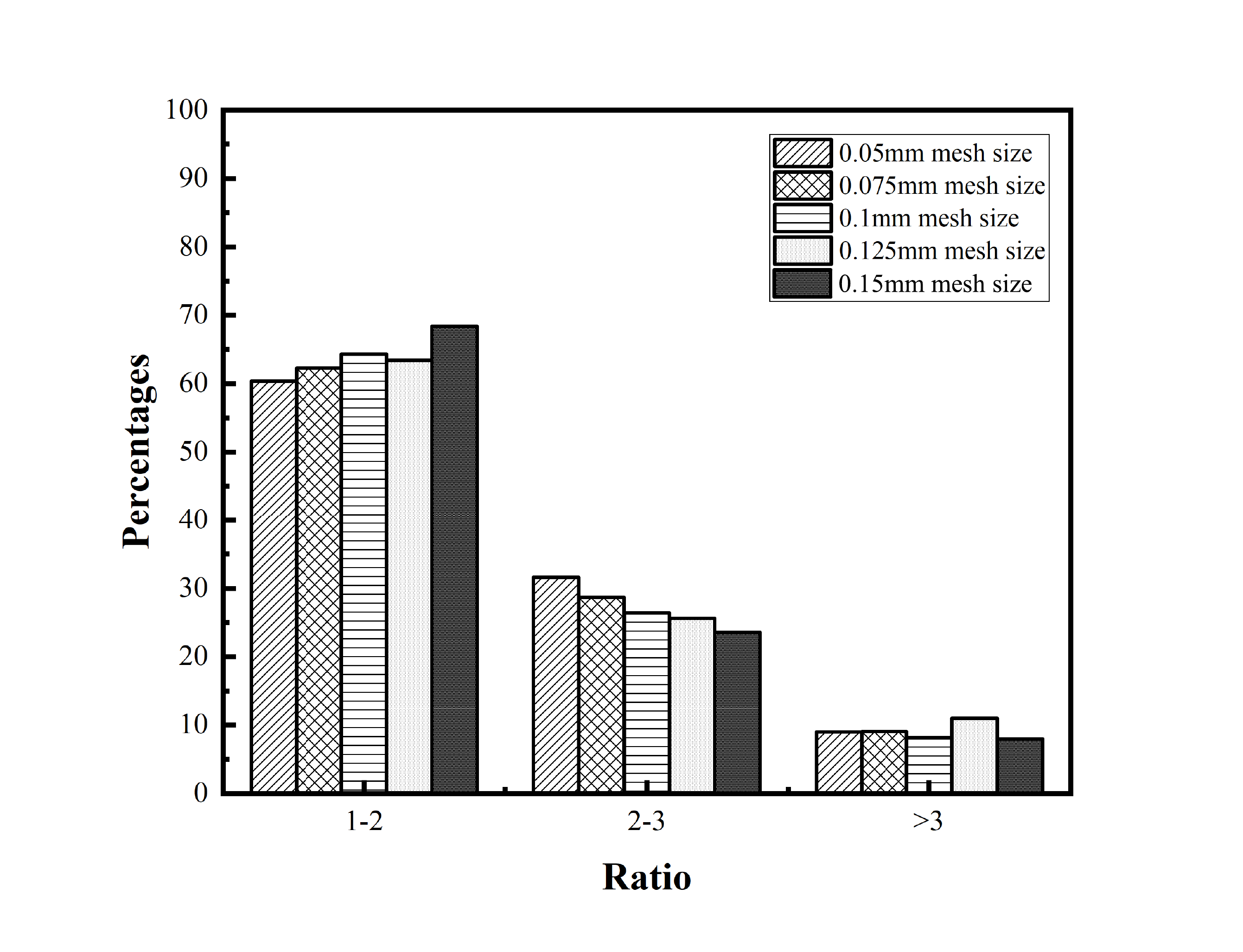}}
\caption{Ratio distribution in four meshes.}
\label{fig_6}
\end{figure}
%----------------------------------------------------------

The whole solution domain is divided into 290,953 nonoverlapping tetrahedrons with 64,765 nodes and 387,700 edges. The overall count of unknowns in the FEM and the ES-FEM are 64,765. The potential along the central axis obtained from the ES-FEM with the linear shape function, the traditional FEM with the linear shape function. The reference solution is obtained from the Comsol with 278,488 unknowns and curved second order shape functions. 

% Fig.7
%----------------------------------------------------------
\begin{figure}[!t]
\centerline{\includegraphics[scale=0.30]{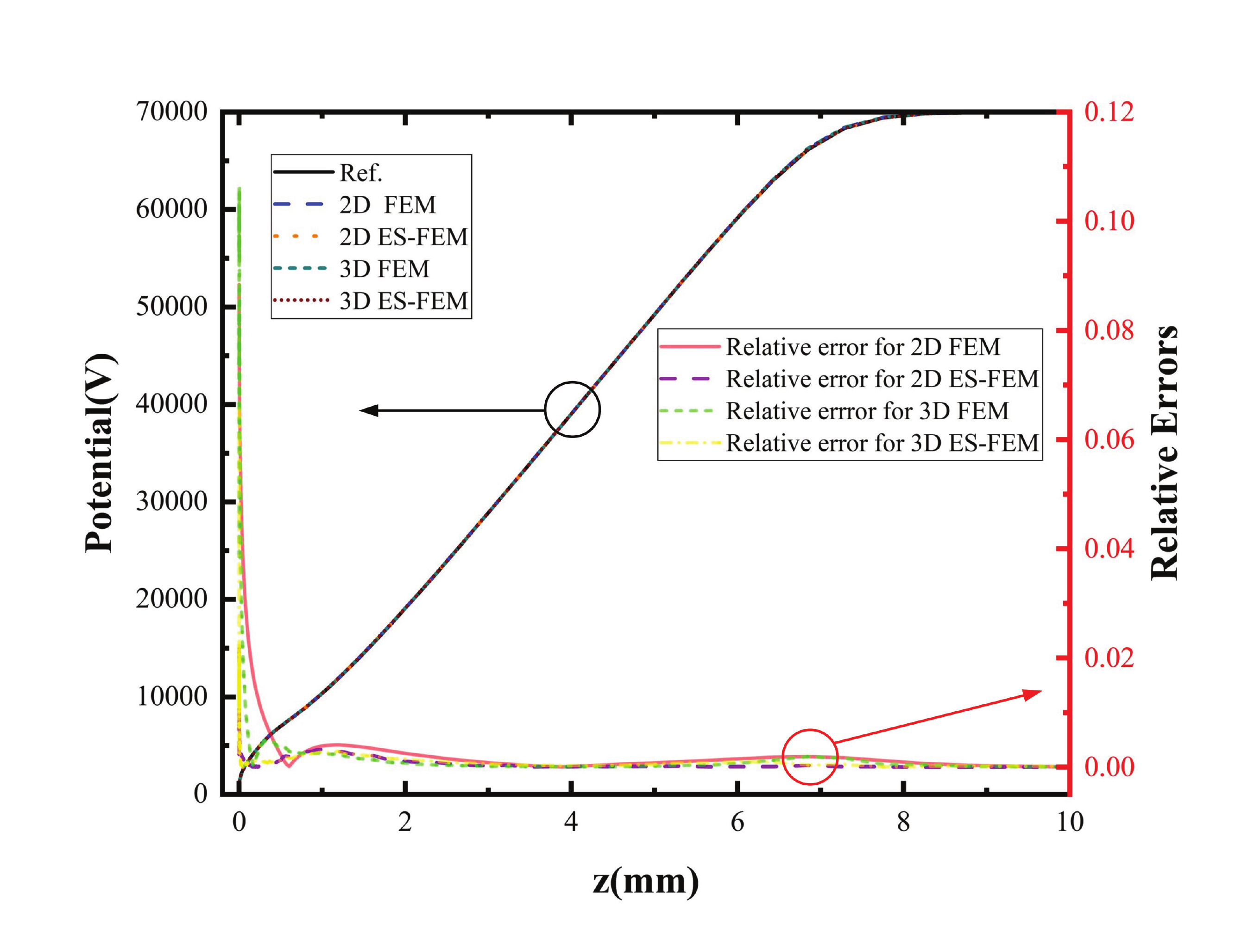}}
\caption{Potential and relative errors of the electronic lens along the axis.}
\label{fig_7}
\end{figure}
%----------------------------------------------------------

% Fig.8
%----------------------------------------------------------
\begin{figure}[!t]
\centerline{\includegraphics[scale=0.30]{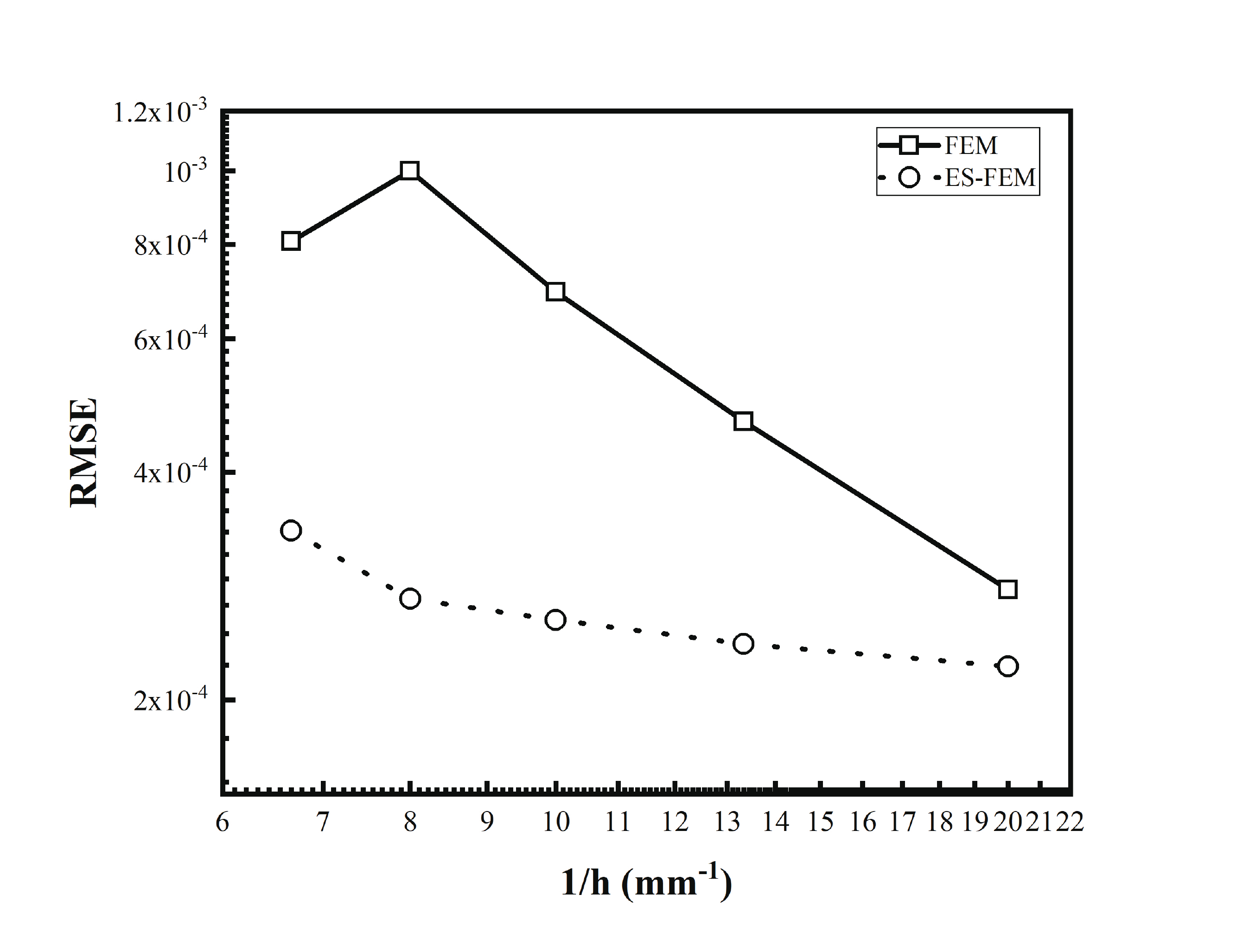}}
\caption{Relative error of the electronic lens with four different meshes.}
\label{fig_8}
\end{figure}
%----------------------------------------------------------

Potential and RMSE along the axis are shown in Fig. 7. Both the two dimensional and three dimensional ES-FEM can obtain much more accurate results, especially in the mesh distortion region, near the tip of the emitter, compared with the FEM. The RMSE of the three dimensional traditional FEM reach 10$\%$, which is almost two times of the three dimensional ES-FEM results and ten times of two dimensinal ES-FEM. The ES-FEM show great numerical stability in the distortion mesh due to the smoothing gradient technique. Furthermore, numerical results show that the ES-FEM in two dimensional system performs better than its three dimensional counterpart due to the geometry symmetry which reduces the error caused by mesh discretization in multiscale structures. 

The relative errors of the three dimensional ES-FEM with four different meshes is shown in Fig. 8. It is easy to find that the ES-FEM is much more accurate than the traditional FEM. It demonstated that the ES-FEM can accurately model multiscale electromagnetic problems.

\subsection{A Micromotor}
We consider a electrostatic micromotor, which is a fully three dimensional structure \cite{ref19}. Due to its symmetrical property, only a quarter of the structure is considered, which has three poles at stator and two poles at the rotor. The radius of stator is 100 $\mu$m with 20 $\mu$m in height. The inner and outer radius of the rotor are 20 $\mu$m and 50 $\mu$m with 18 $\mu$m in height, respectively. The potential of the central stator is set to 100 V and other stators are set to 0 V. Other boundaries are set as Neumann boundary conditions.

% Fig.9
%----------------------------------------------------------
\begin{figure}[!t]
	\begin{minipage}[h]{0.48\linewidth}\label{fig_10A}
		\centerline{\includegraphics[width=1.85in]{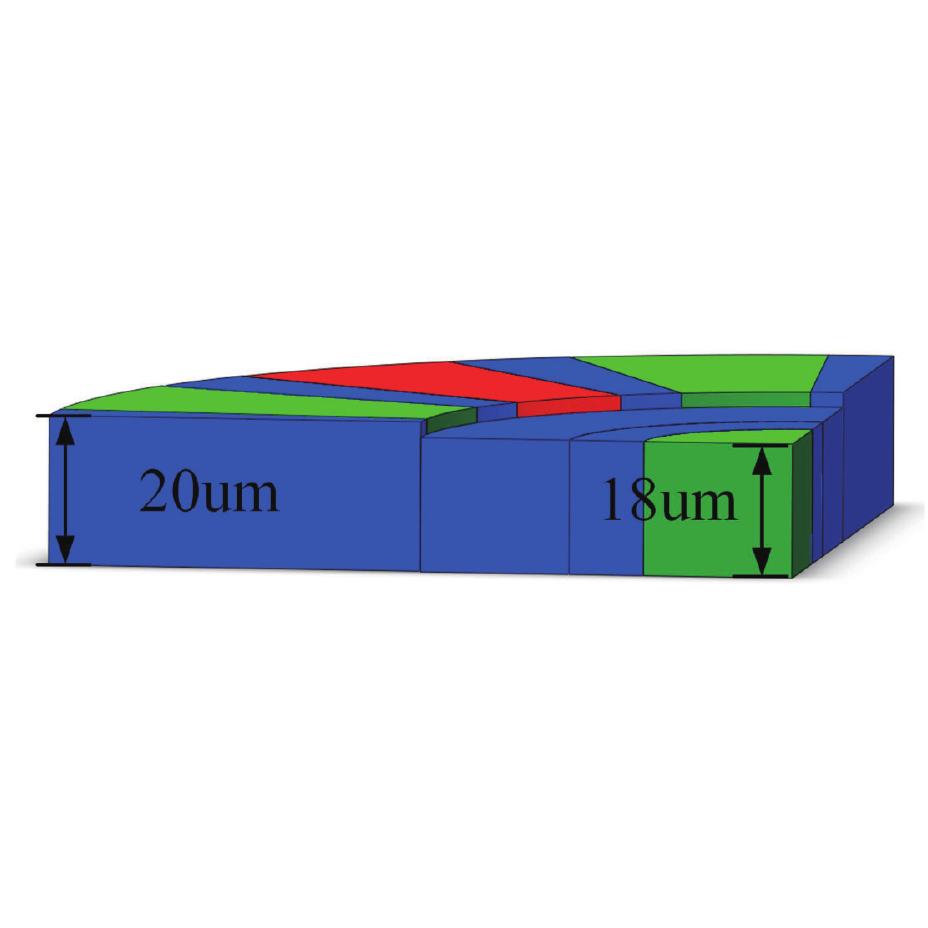}}
		\centerline{(a)}
	\end{minipage}
	\hfill
	\begin{minipage}[h]{0.48\linewidth}\label{fig_10B}
		\centerline{\includegraphics[width=1.8in]{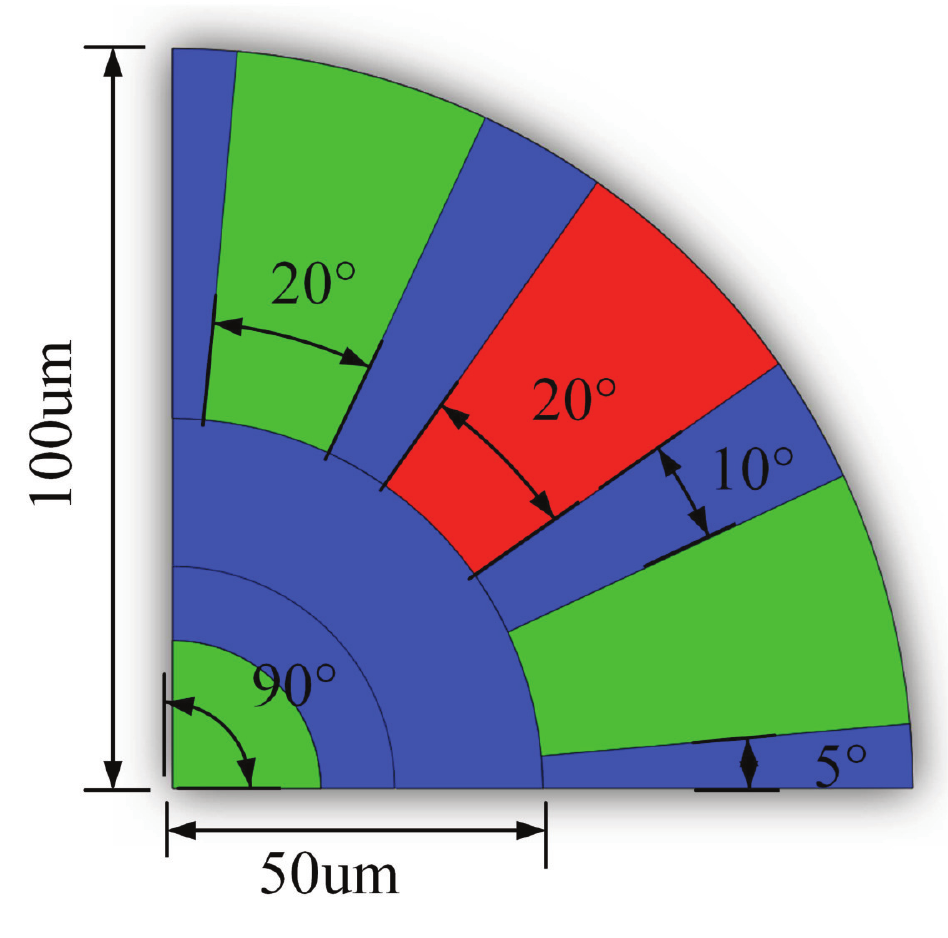}}
		\centerline{(b)}
	\end{minipage}
	\caption{Geometry configurations of the micromotor.}
	\label{fig_9}
\end{figure}
%----------------------------------------------------------
The whole solution domain is discretized into 45,519 nonoverlapping tetrahedrons with 8,839 nodes and 56,815 edges. The potential distribution is computed at $z = 12$ $\mu$m plane as shown in Fig. 10. The potential is also evaluated at $r = 49$ $\mu$m  in the $z = 12$ $\mu$m plane. The reference solution is the numerical solution computed by the FEM with a fine mesh with 509,451 unknowns. As shown in Fig. 10, the ES-FEM with much less unknowns can obtain the exactly the same pattern as the reference. Furthermore, as shown in Fig. 11, the potential obtained from the ES-FEM agrees well with the reference solution. It implies that the ES-FEM can solve complex electromagnetic problems.
% Fig.10
%----------------------------------------------------------

\begin{figure}[!t]
	\begin{minipage}[h]{0.48\linewidth}\label{fig_11A}
		\centerline{\includegraphics[width=2.35in]{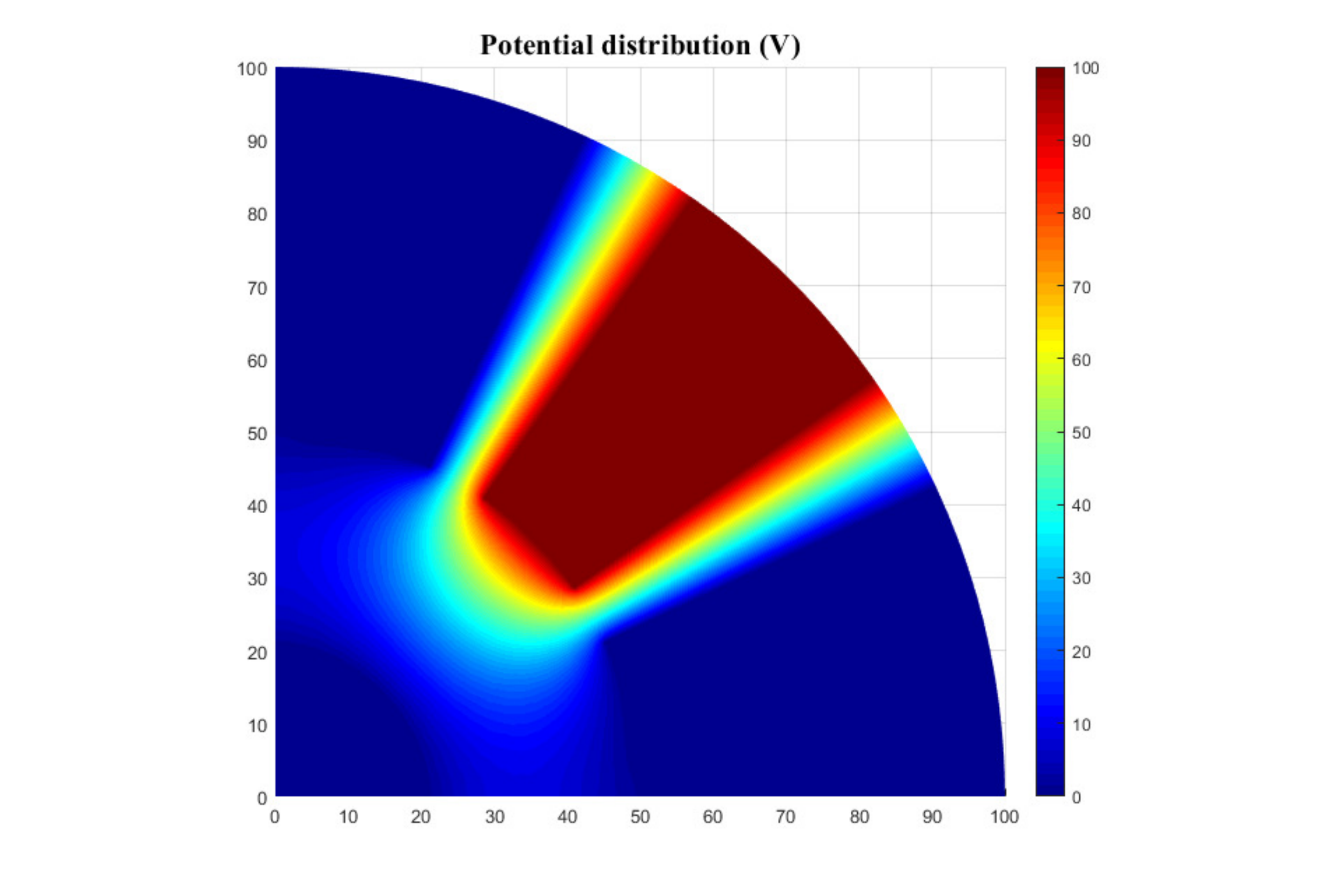}}
		\centerline{(a)}
	\end{minipage}
	\hfill
	\begin{minipage}[h]{0.48\linewidth}\label{fig_11B}
		\centerline{\includegraphics[width=2.35in]{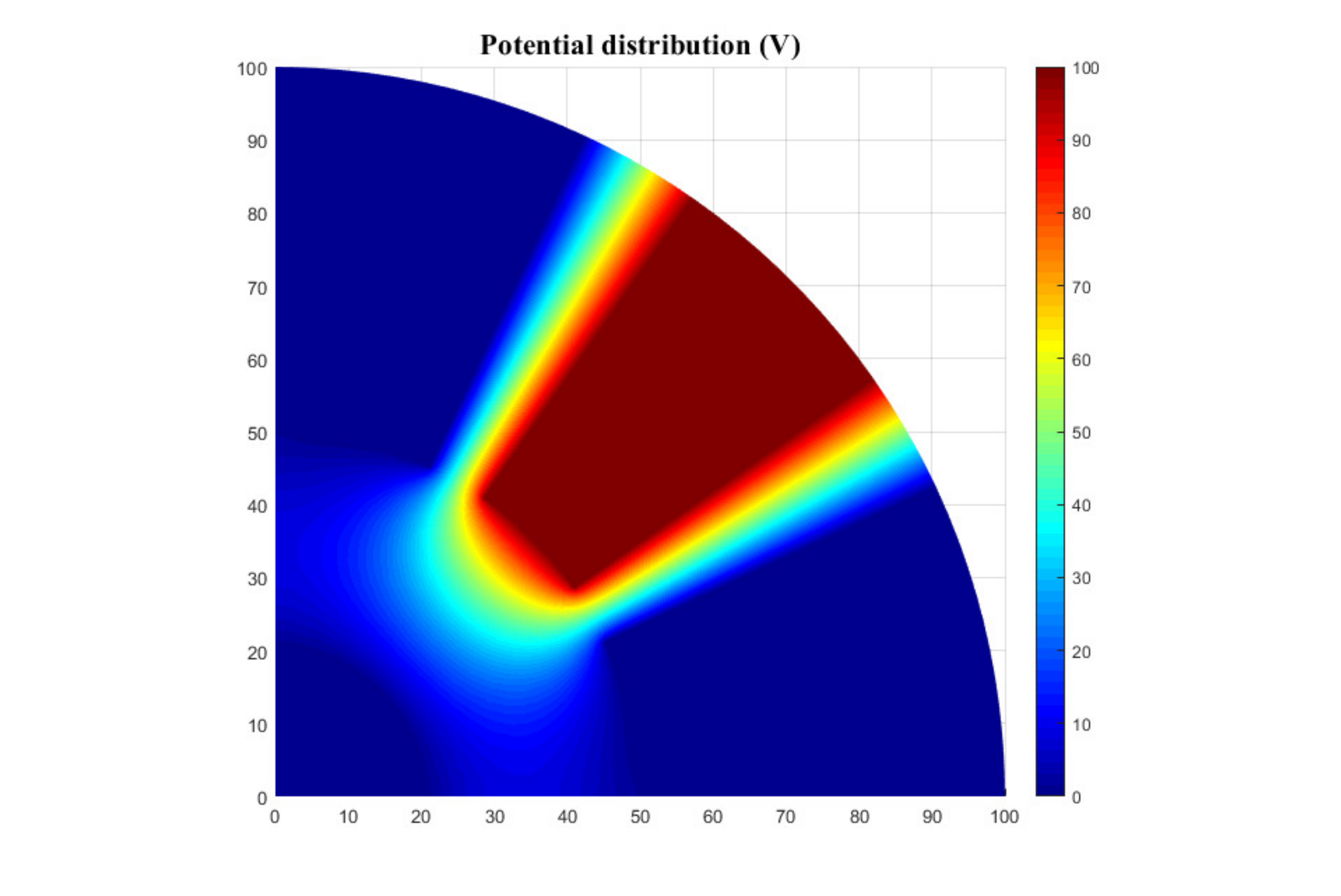}}
		\centerline{(b)}
	\end{minipage}
	\caption{Potential distribution of the micromotor obtained with (a) the FEM with a very fine mesh and (b) the ES-FEM with a coarse mesh. }
	\label{fig_10}
\end{figure}

%----------------------------------------------------------

% Fig.11
%----------------------------------------------------------
\begin{figure}[!t]
\centerline{\includegraphics[scale=0.30]{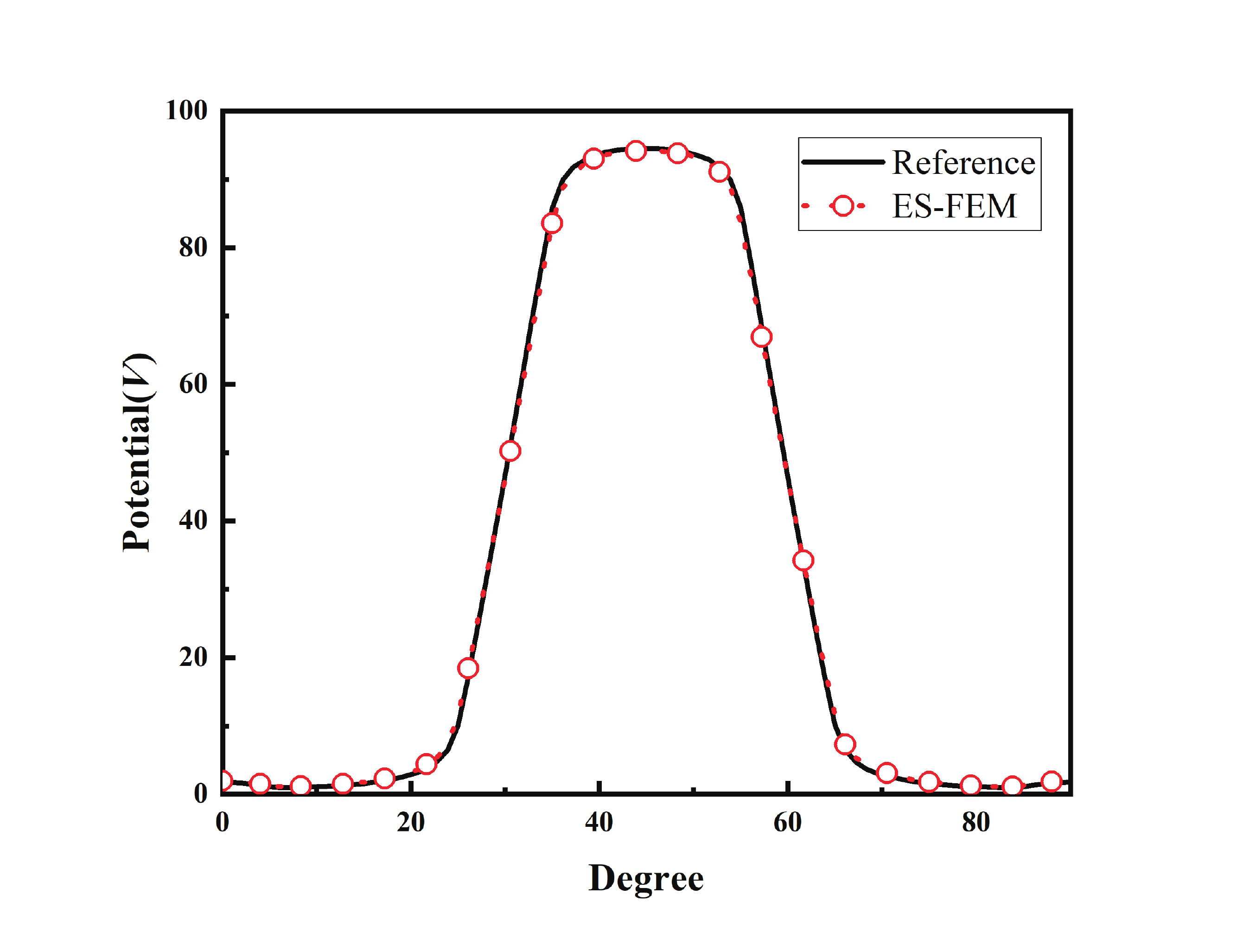}}
\caption{Potential at $z = 12 \mu$m, $r = 49\mu$m obtained from the FEM with a very fine mesh and the ES-FEM with a coarse mesh.}
\label{fig_11}
\end{figure}
%----------------------------------------------------------

\section{Conclusion}
In this paper, the ES-FEM is introduced to accurately solve two dimensional cylindrical and three dimensional electromagnetic problems. By applying the generalized GST to smoothing the gradient component in the FEM and constructing the smoothing domains based on all edges in the background meshes, a highly accurate and numerically stable ES-FEM is constructed. As numerical results show that the ES-FEM can indeed obtain much more accurate results and is stable to irregular meshes, especially when a large number of irregular meshes may exist in the multiscale structures. Therefore, it shows great potential to solve practical electromagnetic problems. Extension of current work to solve full vector electromagnetic problems is in progress. We will report more results upon this topic in the future.

%\section*{Acknowledgment}
%The authors would like to thank financial supports from Beijing Natural Science Foundation through Grant 4194082, National Natural Science Foundation of China through Grant 61801010, and Fundamental Research Funds for the Central Universities.

\end{document}